\documentclass[11pt]{article}
\usepackage{epsfig}

\usepackage{amssymb,amsmath}
\usepackage{epigraph}

\newcommand{\beq}{\begin{equation}}
\newcommand{\eeq}{\end{equation}}
\newcommand{\bea}{\begin{eqnarray}}
\newcommand{\eea}{\end{eqnarray}}

\begin{document}


\begin{flushright}
\end{flushright}
\begin{center}
{\LARGE
Keep It Real:
\vskip0.5cm
Tail Probabilities of Compound Heavy-Tailed Distributions 
} 
\vskip1.0cm
{\Large Igor Halperin} \\
\vskip0.5cm
NYU Tandon School of Engineering \\
\vskip0.5cm
\today \\
\vskip0.5cm 
\vskip0.5cm
{\small e-mail: $igor.halperin@nyu.edu $}
\vskip1.0cm
{\Large Abstract:\\}
\end{center}
\parbox[t]{\textwidth}{
We propose an analytical approach to the 
computation of tail probabilities of compound 
distributions whose individual components have heavy tails. Our approach 
is based on the contour integration method, and gives rise to a representation of the 
tail probability of a compound distribution in the form of a rapidly convergent one-dimensional 
integral involving a discontinuity of the imaginary part of its moment generating function 
across a branch cut. The latter integral 
can be evaluated in quadratures, or alternatively represented as an asymptotic
expansion. Our approach thus offers a viable (especially at high percentile levels) alternative to more standard methods  
such as Monte Carlo or the Fast Fourier Transform, traditionally used for such problems. 
As a practical application, we use our method to compute the operational Value at Risk (VAR) of a financial institution, where individual losses are modeled as spliced distributions whose large 
loss components are given by power-law or lognormal distributions. Finally, we 
briefly discuss extensions of the present formalism for 
calculation of tail probabilities of compound distributions made of compound 
distributions with heavy tails.}



\section{Introduction}

Many practical problems in applied science 
require accurate estimations of tails of compound distributions, i.e. random or non-random sums of random variables. In particular, in the context 
of financial risk management, to
compute the operational Value at Risk (VAR) of a financial institution, the total 
loss resulting from aggregation of losses in 
different lines of business should be calculated at percentile levels of up to 99.9\%. It is well known that "classical" methods 
of calculation of aggregate distributions such as Monte Carlo, Fast Fourier Transform (FFT), or 
the Panjer recursion are
all faced with various numerical issues at such high percentiles. Monte Carlo is robust but slow 
in suppressing the 
simulation noise. Likewise, the FFT and Panjer recursion methods
become slow and lose accuracy, unless some special tricks are used, when computing such 
extreme tails of aggregate loss distributions. 

We propose an analytical approach to calculation of tail probabilities of compound 
distributions where individual components have heavy tails. Our method 
employs a contour integration technique to compute tail probabilities 
expressed in terms of 
moment generating functions (MGFs)  $M(z) $ of corresponding distributions. As is 
well known, in the case of extreme percentile levels such as 
99.9\%, such contour integral representations
of tail probabilities give rise to exponentially small tail probabilities that are formally represented by 
highly oscillating contour integrals. 
We use analyticity in the complex $ z$-plane   
in order to find a suitable deformation of the contour that produces  
a representation of the 
tail probability of  in the form of a rapidly convergent real-valued 
one-dimensional integral. 
Our method is inspired by techniques 
developed for similar problems 
in physics where exponentially small probabilities expressed via highly oscillating 
contour integrals are encountered, in particular, in quantum mechanics \cite{Landau} and 
quantum field theory \cite{ZJ}. 
To the best of our knowledge, such methods were not previously
applied to the problem of computing tail probabilities of compound distributions with heavy tails. 
 
As our procedure produces a representation of tail probabilities in terms of rapidly convergent 
integrals, the latter can be evaluated in quadratures, or alternatively represented as asymptotic
expansions.
Our method thus gives rise to a very efficient numerical scheme
 which avoids numerical issues that arise within
the "classical" methods such as Monte Carlo or FFT when computing such small tail probabilities. 

While the method developed in this paper can be 
applied in many different problems that require accurate computation of tail probabilities for 
single or compound distributions with heavy tails, here we focus on one practical application.
Specifically, we consider the problem of computing the Value at Risk (VAR) for aggregate operational losses of a financial institution. In this setting, the aggregate (compound) operational 
loss of each business unit of the institution can be modeled as    
a compound Poisson process, where each individual loss has a spliced distribution whose two component describe 
small and large losses, respectively. We consider in details two specifications of a single-unit 
large-loss
component: a power-tail (Pareto) or lognormal, however
the same method can 
be applied to other severity distributions as well,  under certain technical conditions 
to be discussed below. 

Furthermore, our analytical formulae assist a model selection 
process by directly capturing the dependence of the 
tail probability on the "body" distribution (i.e. the distribution of small losses). As will be shown below, in models where single loss distributions have heavy tails, VAR at a high percentile level
such as 99.9\% is nearly independent of small losses, where small corrections to the independence 
law depend, for all practical purposes, only on the mean and variance of the distribution of small losses, but not on its higher moments. While the strict asymptotic (i.e. when VAR is yet much higher than 99.9\%) 
decoupling of VAR from small losses for models with heavy tails was established a while ago in Ref.\cite{BK}\footnote{This work was subsequently extended to include the first- and second-order
corrections, see e.g. Ref.~\cite{Sahay} and references therein. Recently, a formal perturbative expansion of tail probabilities for heavy-tailed distributions was proposed by Hernandez et. al. 
\cite{Hernandez}, however their method does not seem to produce a converging or asymptotic expansion.}, our method gives 
an essentially {\it exact} solution that is valid for the actual level of 99.9\% needed in practice, and recovers the result 
of Ref.~\cite{BK} along with all corrections.

\section{A spliced heavy-tailed distribution}

In this paper, we are interested in tail probabilities of compound distributions whose individual components have heavy tails. To have a slightly more general framework, we model individual 
components as spliced distributions with the following probability density function (pdf):
\bea
\label{spliced}
f(x) \, = \, \left\{ \begin{array}{clcr}
(1- \omega) f_1(x) &  \mbox{if $ x \leq x_0$}  \\ 
\omega f_2(x)  &  \mbox{if $ x \geq x_0$} \\  
\end{array} \right. 
\eea  
where both $ f_1(x) $ and $ f_2(x) $ are valid pdf's (which means, in particular, that 
they separately integrate to one). While the second component $ f_2(x) $ has a heavy tail, the first component $ f_1(x) $ can be 
arbitrary. Obviously, the case of a pure heavy-tailed distribution is recovered when $ \omega = 0 $ and 
$ x_0 = 0 $ (assuming that $ x \geq 0 $).

In the specific case of operational risk research that motivated the present work, a spliced 
loss distribution (\ref{spliced}) can be used to model individual losses in a particular unit of measure (UoM)\footnote{Unit of measures is a composition of 
a business line and type of loss that specifies a loss category. 
}, which further compound to produce the total loss in this UoM.
We note that operational losses that belong in the same UoM may be very different 
from each other if they have qualitatively different origins.
For example, for a large investment bank, legal losses can exceed other operational losses by a few orders of magnitude. A spliced loss severity distribution 
(\ref{spliced}), where components $ f_1(x) $ and $ f_2(x) $ might have 
different scales, seems appropriate to model such cases, while reducing to a pure heavy-tailed
distributions in other cases where the use of a spliced distribution is not required. 
For this reason, we stick in this paper to a more general definition of a heavy-tailed distribution as a
spliced distribution (\ref{spliced}). 
Bearing in mind this application of the presented formalism, in what follows we will occasionally refer to the random 
variable $ x $ as the loss severity, though it might correspond to another random variable (e.g. 
the price of a security or a derivative instrument) in other settings.


The two components $ f_1(x) $ and $ f_2(x) $ in Eq.(\ref{spliced})) correspond to 
distributions of small ($ x < x_0 $) and large ($ x \geq x_0 $) losses, respectively,
where $ x_0 $ is a right tail threshold. 
The mixing 
parameter $ 0 \leq \omega \leq 1 $ can be chosen e.g. by the continuity condition
\beq
\label{junction}
(1 - \omega) f_1(x_0) = \omega f_2(x_0)
\eeq
Alternatively, we can integrate Eq.(\ref{spliced}) from $ 0 $ to $ x_0 $ and replace the unknown 
cumulative distribution $ F(x) = \int_{0}^{x} f(s) ds $ by the empirical distribution. This provides 
a model-independent relation\footnote{We refer to (\ref{omega}) as a 
model-independent relation as, according to the Glivenko-Cantelli theorem, 
the empirical distribution 
$ F_{emp}(x) = \frac{1}{N}\sum_{i=1}^{N} \theta(X_i \leq x) $ converges to the 
true distribution $ F(x) $ uniformly 
as $ N \rightarrow \infty $ almost surely.}
\beq
\label{omega}
\omega = 1 - F_{emp}(x_0)
\eeq
Note that if we fix $ \omega $ using Eq.(\ref{omega}), then Eq.(\ref{junction}) can be 
considered as a constraint on the low-loss distribution $ f_1(x) $ at the junction 
point $ x_0 $ (provided we know distribution $ f_2(x) $), rather
than a condition on $ \omega $.  
In what follows, we do not dwell further on the modeling of $ f_1(x) $ and 
keep it largely unspecified because, as 
will be shown
below, VAR figures are  
mostly driven by the second, high-loss component $ f_2(x) $. The dependence 
on $ f_1(x) $
comes only through corrections in an asymptotic expansion in powers of $ 1/s $ 
(where $ s $ is the VAR loss level), 
which are expected to be fairly small for sufficiently 
high percentile levels such as 99.9\%.   

In this paper, we consider two specifications of the severity distribution: a power-law and a 
lognormal law. Our choice is motivated by the observation that these 
two distributions seem to 
provide the best fit (of about equal quality) to large losses for the majority 
of UoMs of large financial institutions.
In the next section, we concentrate on a power-law distribution, while the case of a lognormal distribution 
will be analyzed in Sect.4.
 
 We consider the following power-law distribution for
 losses above the threshold $ x_{0} $:
\beq
\label{power_law}
f_2(x) = f(x|x \geq x_{0}) = \frac{\alpha - 1}{x_{0}} \left( \frac{x}{x_{0}} \right)^{- \alpha} \equiv C x^{ - \alpha}
\eeq
where  $ C = (\alpha-1) x_{0}^{\alpha - 1} $ is the 
normalization constant. The tail distribution is 
\beq
\label{ccdf_power}
\bar{F}_2(x) = \int_{x}^{\infty} ds \, f_2(x) = \left( \frac{x}{x_{0}} \right)^{- (\alpha -1)}
\eeq 
The constant $ \alpha > 1 $ in (\ref{power_law}), (\ref{ccdf_power}) 
is called the {\it exponent} of the power law.
Note that, given a choice of $ x_{0} $, the exponent $ \alpha $ is the {\it only} free parameter
in the power-law distribution.  
The fitting procedures to compute parameters $ \alpha $
and $ x_{0} $ will be described below. For now, we note that the optimal value of 
$ \alpha $ depends on $ x_{0} $ logarithmically (i.e. mildly). 

Power-law distributions (also known as Zipf or Pareto distributions) are ubiquitous in both 
nature and society,
see e.g. \cite{Newman_2006} for a review, and are often characteristic of complex systems. Ref. \cite{Newman_2006} gives many examples of power-law distributions
in language, demography, commerce, computer science, information theory, physics, astronomy,
geology and so on. It seems that among "wide" distributions (i.e. distributions for random variables 
that may vary by
several orders of magnitude), a power-law distribution is more often a rule rather than an exception. 
For wide distributions that are {\it not} power-law, 
Ref. \cite{Newman_2006} mentions some (not too many) examples that are better described by a 
log-normal or a "stretched exponential" (Weibull) distribution.

In our experiments with real-world operational loss datasets, we 
found that the power-law distribution (\ref{power_law}) outperforms the Weibull 
and log-Weibull distributions in terms of both stability and quality of 
matching the data in the tail, while performing about equally well with a lognormal distribution.  

\section{Tail probabilities by contour integration}

\subsection{Moment generating functions}

We start with the moment-generation function (MGF) of the power-law (Pareto) distribution:
\beq
\label{MGF_0}  
M(z) = \mathbb{E} \left[ e^{-zX} \right] = (\alpha- 1) \left(z x_{0} \right)^{\alpha-1} 
\Gamma \left(1-\alpha, z x_{0} \right) \, , \; \; z \geq 0
\eeq
Note that our sign convention is such that the MGF defined by (\ref{MGF_0}) coincides 
with the Laplace transform of the distribution\footnote{It is more common
to define the MGF as $  \mathbb{E} \left[ e^{zX} \right] $.}.  
In what follows, we will also use the characteristic function (CF)
of $ f(x)$ which is equal to the MGF evaluated at a purely imaginary argument:
\beq
\label{CF}
\phi(z) = \mathbb{E} \left[ e^{izX} \right] = M(-iz)
\eeq
Note that the upper incomplete gamma function $ \Gamma \left(s, z \right) $ 
has the following asymptotic behavior:
\bea
\label{asympt_Gamma}
\Gamma(s,z) \, = \, \left\{ \begin{array}{clcr}
z^{s-1} e^{-z}  \left[ 1 + \frac{s-1}{z} + \ldots \right]
,  &  
\mbox{if $ | z| \rightarrow \infty, \, | \arg \, z |   < \frac{3}{2} \pi $}  \\ 
\Gamma(s) -  z^s \left[ \frac{1}{s} - \frac{z}{s+1} + \ldots \right], 
&  \mbox{if $ | z| \rightarrow 0, \,  Re \, s <  0$} \\  
\end{array} \right. 
\eea 
This implies the following behavior of $ M(z) $ in the limits $ z \rightarrow 0 $ and 
$ |z| \rightarrow \infty $:
\bea
\label{asympt_M}
M(z) \, = \, \left\{ \begin{array}{clcr}
\frac{1}{x_0 z} e^{-x_0 z}  \left[ 1  -  \frac{\alpha}{x_0 z} + O(z^{-2}) \right]
,  &  
\mbox{if $ | z| \rightarrow \infty, \, | \arg \, z |   < \frac{3}{2} \pi $}  \\ 
1 - \Gamma(2-\alpha) (x_0 z)^{\alpha-1} + \frac{\alpha-1}{2-\alpha} x_0 z + O(z^2), 
&  \mbox{if $ | z| \rightarrow 0, \, \alpha > 1 $} \\  
\end{array} \right. 
\eea  
The upper incomplete gamma function $ \Gamma \left(s, z \right) $ 
can be expressed in terms of the confluent hypergeometric function, also 
known as Kummer's function\footnote{Other notations 
for this function are $ M(a,b,z)$ and $ _{1}F_{1}(a;b;z) $.}:
\beq
\label{hypergeom}
\Gamma(s, z) = \Gamma(s) - \frac{z^{s}}{s}  \Phi(s, 1+s, -z)
\eeq
Using this in (\ref{MGF_0}), we obtain
\beq
\label{MGF}
M(z) = \Phi(1-\alpha, 2 - \alpha,  -x_0 z) - \Gamma(2 - \alpha) \left( x_0 z \right)^{\alpha-1}
\eeq
Note that Kummer's function $ \Phi(a, b, z) $ is 
an {\it analytic} (holomorphic) function of $ z $, while a power function in the 
second term in (\ref{MGF}) has a {\it branch cut singularity} 
in the complex $ z $-plane, which can 
be chosen to be  the interval $ [- \infty, 0] $. 
We note that other heavy-tailed distributions, including in particular a lognormal distribution,
have MGFs with similar analytic properties. 
Our approach developed below is very general and applies to any distribution 
whose MGF is analytic in a 
complex plane with a branch cut singularity. While the explicit form of the MGF is not used,
the only additional requirement for the method to be applicable is that the MGF does not grow 
too fast in the left semi-plane. In particular, the MGF of the Pareto distribution is well behaved in
this sense. Indeed, as shown in Eq.(\ref{asympt_M}), $ M(z) $ grows asymptotically $ \sim 
e^{-x_0 z} $ in the left semi-plane, but this divergence is integrable when calculating the tail probability
(see below). 

For the spliced distribution (\ref{spliced}), the MGF function has a similar form to Eq.(\ref{MGF}):
\beq
\label{MGF_s}
M(z) = R(z) - \omega \Gamma(2 - \alpha) \left( x_0 z \right)^{\alpha-1}
\eeq
where we defined function $ R(z) $ as follows:
\beq
\label{R}
R(z) = \omega \Phi(1-\alpha, 2 - \alpha,  -x_0 z)  + (1-\omega) M_1(z)
\eeq
where $ M_1(z) $ stands for the MGF of the "body" distribution $ f_1(x) $. We assume that this 
distribution has all moments. In this case, $ M_1(z) $ is an analytic function of $ z $, and therefore 
function $ R(z)$ is analytic as well. Note that $ R(0) = 1 $.

To compute the MGF of a compound distribution, we need to specify a loss frequency model.
For simplicity, we assume that the loss frequency for the time horizon $ T = 1 $ is given by 
a Poisson distribution with intensity $ \lambda $. The loss severity and loss frequency processes 
are assumed to be independent. Individual losses are independent as well. In this case, 
the MGF $ M_{\lambda}(z) $ 
of the compound process (a random sum $ X_1 + \ldots + X_n $ of individual losses
where $ n $ is randomly drawn from the Poisson distribution) can be computed explicitly:
\beq
\label{MGF_c}
M_{\lambda} (z) = \sum_{n=0}^{\infty} \frac{(\lambda T)^n}{n!} e^{ - \lambda T} \left[ M(z) \right]^{n} 
= e^{ \lambda T ( M(z) - 1)}
\eeq
 Note that as $ e^{z} $ is an entire function (i.e. it is analytic in the whole complex plane), the compound MGF $ M_{\lambda}(z) $ is 
 analytic in the complex $ z$-plane with the same branch cut singularity as 
 the one present for the single loss MGF $ M(z) $.
 
 \subsection{Tail probability as a contour integral}

Recall the integral representation of the Heaviside step-function $ \theta(x) $:
\beq
\label{Heavi}
\theta(x) = \frac{e^{x \varepsilon}}{2 \pi i } \int_{- \infty}^{\infty} \frac{ e^{izx}}{z - i \varepsilon} dz
\eeq
where $ \varepsilon > 0 $ is arbitrary. In practice, it is convenient to employ the limit $ \varepsilon
\rightarrow +0 $, which is what will be assumed in what follows. Using the residue theorem, it is easy
to check that $ \theta(x)$ defined by Eq.(\ref{Heavi}) is one when $ x > 0 $, and zero when $ x < 0 $.
Indeed, when $ x > 0 $, we close the contour in the upper semi-plane, and the integral is one due to the residue at the pole at $ z = i \varepsilon $. Otherwise, if $ x < 0 $, we close the contour in the lower
semi-plane, and the integral equals zero.

\begin{figure}[ht]
\begin{center}
\includegraphics[width=61.92mm,height=74.04mm]{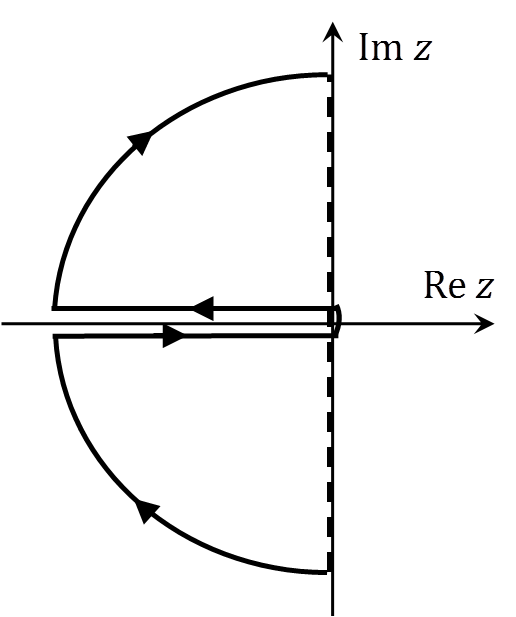}
\caption{Integration contour for integral (\ref{tail_3}).} 
\label{fig:semicircle}
\end{center}
\end{figure}

Using Eq.(\ref{Heavi}), we can write the tail probability of a loss distribution with pdf $ p(x)$ as 
follows:
\beq
\label{tail_one}
\bar{F}(s) \equiv P \left( X > s \right) = \int_{0}^{\infty} dx \, p(x) \theta(x - s) =  \int_{0}^{\infty} dx \, p(x)
\frac{e^{(x-s) \varepsilon}}{2 \pi i } \int_{- \infty}^{\infty} \frac{ e^{iu(x-s)}}{u - i \varepsilon} du
\eeq
Exchanging the order of two integrations, we obtain
\beq
\label{tail_two}
\bar{F}(s) =   \frac{e^{-s \varepsilon}}{2 \pi i }  \int_{- \infty}^{\infty} \frac{ e^{- i s u}}{u - i \varepsilon} \phi( u - i \varepsilon) du
\eeq
where $ \phi(z)$ stands for the characteristic function of distribution $ p(x) $:
\beq
\label{CF_p}
\phi(z) = \mathbb{E} \left[ e^{ i z X} \right] = \int_{0}^{\infty} dx \,  e^{ i z x}  p(x)
\eeq 
Introducing a new variable $ z $ by the relation $ u = i z + i \varepsilon $, we obtain
\beq
\label{tail_3}
\bar{F}(s) =  -\frac{1}{2 \pi i }  \int_{- \varepsilon - i \infty}^{- \varepsilon + i\infty} 
\frac{ e^{s z}}{z} \phi( i z) dz = - \frac{1}{2 \pi i }  \int_{- \varepsilon - i \infty}^{- \varepsilon + i\infty} 
\frac{ e^{s z}}{z} M(z) dz
\eeq 
Note that contour of integration in Eq.(\ref{tail_3}) runs parallel to the imaginary axis. 
We assume that the MGF $ M(z) $ is such that $ e^{sz} M(z)  \rightarrow 0 $ for 
sufficiently large $ s $ when 
$ |z| \rightarrow \infty $ with $ Re \, z < 0 $ (we will return to this point below when we consider
specific applications). In this case, 
we can produce a closed contour by adding a semi-circle $ C: 
|z| = R \rightarrow \infty , Re \, z < 0  $ in the left semi-plane to our initial open contour.
By Jordan's lemma, the value of the integral over the closed contour is equal to the value of the 
original integral with the open contour.
Once we have a closed contour, we can use the Cauchy integral theorem and 
arbitrarily deform it within the analyticity domain without changing the value of the integral.
As 
$ M(z) $ is analytic in the complex $ z$-plane with a branch cut on $ z \in [-\infty, 0] $, we 
can "squeeze" the integration contour such that it runs first under the cut for $ - \infty $ to $ 0 $, 
then flips around the origin $z = 0 $, and runs back to $ - \infty $ above the cut:
\beq
\label{tail_4}
\bar{F}(s) = - \frac{1}{2 \pi i} \int_{- \infty}^{0} dz \frac{ e^{s z}}{z} M^{-}(z) - 
\frac{1}{2 \pi i} \int_{0}^{- \infty} dz \frac{ e^{s z}}{z} M^{+}(z)
\eeq
where $ M^{+}(z) $ and $ M^{-}(z) $ stand for the values of the MGF $ M(z) $ 
above and below the 
branch cut, respectively (see Fig.~\ref{fig:semicircle}).

The two integrals in Eq.(\ref{tail_4}) do not cancel out due to a discontinuity of 
the imaginary part of $ M(z)$ across the branch cut at 
$ z = [-\infty,0] $\footnote{The real part of $M(z) $ should be 
continuous across the branch cut as the tail probability should be real.}. 
Setting $ z = x e^{ i \pi} $ and $ z = x e^{-i \pi} $ with $ x \geq 0 $ 
at the upper and lower bank of the cut, respectively, the discontinuity is 
computed as follows:
\beq
\label{Disc_0}
\Delta \mathrm{Im} \,\, M(x) = \mathrm{Im} \,\, M(x e^{  i \pi}) - \mathrm{Im} \,\,  M(x e^{ -i \pi}) 
\eeq 
Using this in Eq.(\ref{tail_4}), we obtain 
\beq
\label{main}
\bar{F}(s) = - \frac{1}{2 \pi } \int_{0}^{\infty} dx \frac{ e^{- s x}}{x} \Delta  \mathrm{Im} \,\, M(x) 
\eeq 
Eq.(\ref{main}) constitutes our first main result. In the 
original complex-valued contour integral (\ref{tail_3}), the 
integrand becomes a strongly oscillating function when
$ s \rightarrow \infty $, which makes it difficult to accurately compute the tail probability 
in this limit. On the other hand, it is exactly this limit that is relevant for computing VAR at
high percentile levels such as 99.9\%.
Using analyticity of the MGF in the complex plane with a cut,   
we have managed to reduce the complex integral 
(\ref{tail_3})  
to a rapidly convergent  integral (\ref{main}) 
defined on the real axis. The latter integral can be computed very 
efficiently and accurately in quadratures. Clearly, numerical integration 
is much faster and more accurate than either Monte 
Carlo or a convolution method which are typically subject to a substantial 
numerical noise 
when computing tail probabilities at high percentiles. 
Using analyticity and contour integration, such 
numerical noise is filtered out in Eq.(\ref{main}). 

Note that the relation (\ref{main}) is very general and 
applies to {\it any} distribution whose MGF $ M(z) $ is analytic in a cut $ z$-plane 
and well-behaved at infinity $ |z| \rightarrow \infty, \, Re \, z < 0 $. 
It does {\it not} apply to distributions whose moment generating functions do not have a branch
cut singularity, as in the letter case the discontinuity of the imaginary part of $ M(z)$ would 
vanish\footnote{For distribution with both a branch cut singularity and additional singularities in the complex plane such as single poles, the present formalism could be extended by taking a proper, 
problem-specific care of these additional singularities.}. We will next consider uses of 
Eq.(\ref{main}) to compute tail probabilities of both the single 
loss and compound loss distributions.

 
\subsection{Tail probability of a single power-law distribution}

As a check of our general relation (\ref{main}), we apply it to compute the tail probability
$ P(X>s) $ of the single loss distribution given by Eq.(\ref{spliced}) where
the second component $ f_2(x) $ is chosen to be a power-law distribution
(\ref{power_law}). Clearly, the answer in this case
can be obtained by elementary means, and reads $ P(X>s) = \omega \bar{F}_2(s) $ where 
$ \bar{F}_2(x) $ is given by Eq.(\ref{ccdf_power}). Therefore, Eq.(\ref{main})
should produce the same result. 
 
 Before proceeding with the calculation, we note that the MGF of a power-law distribution
 (or a distribution with a power-law tail as our Eq.(\ref{spliced})) is well behaved at infinity, 
 as $ e^{sz} M(z)  \rightarrow 0 $ when $ s > x_0 $ and 
$ |z| \rightarrow \infty $ with $ Re \, z < 0 $, as can be seen from Eq.(\ref{asympt_M}). 
Therefore, Eq.(\ref{main}) can be used in this case.

To compute the discontinuity of the MGF (\ref{MGF_s}) across the branch cut, we note that 
it arises solely due to the second term 
in (\ref{MGF_s}). Let $ z = x e^{i \theta} $ along the cut, such 
that the path above the cut corresponds to $ \theta =  \pi $, and the path  below the cut has
$ \theta  = - \pi $. On these paths, the (main branch of the) power function 
$ ( x_0 z)^{\alpha-1} $ takes values $ - ( x_0  x)^{\alpha-1} e^{ i \pi \alpha} $ and
$ - ( x_0  x)^{\alpha-1} e^{- i \pi \alpha} $, respectively.
Therefore, the discontinuity of $\mathrm{Im} \,\, M(z)$ across the branch cut at $ [- \infty,0] $ 
reads\footnote{Note that as $ Re \, e^{ i \pi \alpha} = \cos(\pi \alpha) $ is an even function, 
the real part of $ M(z) $ does not have a discontinuity across the branch cut.}
\bea
\label{disc_1}
\Delta \mathrm{Im} \,\, M(x) &=& \omega 
\Gamma(2- \alpha) (x_0 x)^{\alpha-1} \mathrm{Im} \,\left( e^{ i \pi \alpha} -  
e^{- i \pi \alpha} \right) =  2 \omega \Gamma(2 - \alpha) \sin( \pi \alpha) (x_0 x )^{\alpha-1}
\nonumber \\
&=& - \frac{2 \omega \pi}{\Gamma(\alpha-1)} (x_0 x )^{\alpha-1}
\eea
where at the last step we used the identity
\beq
\label{gamma_ident}
\Gamma(x) \Gamma(1-x) = \frac{\pi}{\sin( \pi x)}
\eeq
Introducing a new variable $ y = x_0 x $ and denoting $ \hat{s} = s/x_0 $, we 
obtain for the integral (\ref{main}):
\beq
\label{main_tail_1}
\bar{F}(s) = 
\frac{\omega}{\Gamma(\alpha-1)}
\int_{0}^{\infty} dy \frac{ 
e^{- \hat{s} y}}{y} y^{\alpha-1}  
=  \omega \left( \frac{s}{x_0} \right)^{-(\alpha-1)} 
\eeq
We have thus verified that Eq.(\ref{main}) correctly reproduces the tail probability for a single 
loss distribution with a power-law tail.

\subsection{Tail probability of a compound power-law distribution}

Now we turn to a more interesting application of relation (\ref{main}). Namely, 
we use it to compute the tail probability of the compound distribution whose MGF is given
by Eq.(\ref{MGF_c}). No explicit closed form answer for this quantity is available, however
we can compare our results with asymptotic expressions in the limit $ s \rightarrow \infty $ 
that are available in the literature, as well as verify our results numerically.

Unlike the previous single-loss case, Eq.(\ref{main}) cannot be straightforwardly used to 
compute the tail probability of the compound distribution, as the MGF (\ref{MGF_c}) 
grows as an exponent of an exponent in the left semi-plane
as implied by Eq.(\ref{asympt_M}). However, this problem is easy to fix. To this end, let us 
identically represent the MGF (\ref{MGF_c}) in the following form:
\bea
\label{MGF_c_2}
M_{\lambda} (z) &=&  e^{ \lambda T ( M(z) - 1)} = 
\sum_{n=0}^{n_0} \frac{(\lambda T)^n}{n!} e^{ - \lambda T} \left[ M(z) \right]^{n}
+  \sum_{n=n_0+1}^{\infty} \frac{(\lambda T)^n}{n!} e^{ - \lambda T} \left[ M(z) \right]^{n} \nonumber \\
&=& \left( M_{\lambda} (z)  
- \mathcal{M}_{\lambda}(z) \right) + \mathcal{M}_{\lambda}(z) \equiv \tilde{M}_{\lambda}(z) + \mathcal{M}_{\lambda}(z)
\eea
where $ n_0 = \lfloor \frac{s}{x_0} \rfloor $ is the largest integer that is smaller or equal to 
the ratio $ \frac{s}{x_0} $, and function $ \mathcal{M}_{\lambda}(z) $ is a Taylor tail of $ M_{\lambda}(z) $: 
\beq
\label{mz}
\mathcal{M}_{\lambda}(z) = \sum_{n=n_0+1}^{\infty} \frac{(\lambda T)^n}{n!} e^{ - \lambda T} \left[ M(z) \right]^{n}
\eeq 
while function $ \tilde{M}_{\lambda}(z) $ is equal to $ M_{\lambda}(z) $ with its 
Taylor tail subtracted. Now the first term $ \tilde{M}_{\lambda}(z) $ in Eq.(\ref{MGF_c_2}) 
is well-behaved in the left semi-plane when $ s \geq x_0 $, while 
the second term $ \mathcal{M}_{\lambda}(z) $ is well-behaved in the right 
semi-plane where the product $ e^{sz} \mathcal{M}_{\lambda}(z)  \rightarrow 0 $ as $ z 
\rightarrow \infty $ when $ s \geq x_0 $. 

Using Eq.(\ref{MGF_c_2}), we can write the contour integral as follows:
\beq
\label{tail_sum}
\bar{F}(s) =  - \frac{1}{2 \pi i }  \int_{- \varepsilon - i \infty}^{- \varepsilon + i\infty} 
\frac{ e^{s z}}{z} \tilde{M}_{\lambda}(z) dz - \frac{1}{2 \pi i }  \int_{- \varepsilon - i \infty}^{- \varepsilon + i\infty} 
\frac{ e^{s z}}{z} \mathcal{M}_{\lambda}(z) dz 
\eeq 
We close the contour for the first integral in the left semi-plane and deform it following steps that led to 
Eq.(\ref{main}), while for the second integral we close the contour in the right semi-plane and apply 
the residue theorem. We obtain\footnote{Note that our method based on using Eq.(\ref{MGF_c_2})
to decompose a formally diverging contour integral into an exactly calculable contribution plus a 
converging term that can be evaluated as an asymptotic expansion is somewhat similar to the celebrated 
Lugannani-Rice method \cite{LR} where a similar trick is used to handle an apparent singularity at 
$ z = 0 $.} 
\beq
\label{main_c}
\bar{F}(s) = - \frac{1}{2 \pi } \int_{0}^{\infty} dx \frac{ e^{- s x}}{x} 
\Delta  \mathrm{Im} \,\, \tilde{M}_{\lambda}(x) + \mathcal{M}_{\lambda}(0)
\eeq 
Here 
\beq
\label{m_0}
\mathcal{M}_{\lambda}(0) = \sum_{n=n_0+1}^{\infty} \frac{(\lambda T)^n}{n!} e^{ - \lambda T} = 
P (X > n_0) 
\eeq
where $ X \sim Po(\lambda) $ stands for a Poisson random variable driven by the Poisson law with intensity $ \lambda T $. Thus $ \mathcal{M}_{\lambda}(0) $ is the tail probability of the Poisson distribution. It 
can be estimated using the large deviation theory
(see Appendix A for details):
\beq
\label{Cramer_tail}
\mathcal{M}_{\lambda}(0) = P \left(  Po(\lambda T) > n_0 \right) \sim  
e^{ - n_0 \left( \lambda T - \log \lambda T - 1 \right) }
\eeq 
Recalling that $ n_0 = \lfloor \frac{s}{x_0} \rfloor $, we see that the $ \mathcal{M}_{\lambda}(0) $ term 
in Eq.(\ref{main_c}) is exponentially suppressed for large $ s $.

Next we turn to the first term in Eq.(\ref{main_c}). We have
\beq
\label{M_lam_tilde}
\tilde{M}_{\lambda}(z) = M_{\lambda}(z) - \mathcal{M}_{\lambda}(z) = 
e^{ \lambda T( R(z) - 1) }  e^{- \omega \lambda T 
\Gamma(2 - \alpha) \left( x_0 z \right)^{\alpha-1} }  - \mathcal{M}_{\lambda}(z)
\eeq
We start with computing the discontinuity of the imaginary part of $ M_{\lambda}(z) $. 
As function $ R(z) $ is analytic, the first exponent in (\ref{M_lam_tilde}) is continuous across 
the branch cut, and discontinuity across the cut is due to the second exponent. Defining 
the values of the power function on different sides of the branch cut in the same way as was done 
in the previous section, we find the discontinuity of the imaginary part of $ M_{\lambda}(z) $:
\bea
\label{disc_comp}
\Delta \mathrm{Im} \,\, M_{\lambda} (x) &=&  e^{ \lambda T( R(-x) - 1) } \Delta \mathrm{Im} \,\, 
\left( e^{\omega \lambda T \Gamma(2-\alpha) (x_0 x)^{\alpha-1} e^{i \pi \alpha}} \right) \\
& = &
- 2 e^{ \lambda T \Psi(x) } \sin \left[ \frac{ \pi \omega \lambda T}{\Gamma(\alpha-1)} 
\left( x_0 x \right)^{\alpha-1} \right] 
\nonumber
\eea
where for convenience we introduced function $ \Psi(x) $ as follows:
\beq
\label{Psi}
\Psi(x) = R(-x) + \omega \Gamma(2 - \alpha) \cos (\pi \alpha) \left( x_0 x \right)^{\alpha-1} - 1 
\eeq
Let us now consider the discontinuity of $ \mathrm{Im} \,\, \mathcal{M}_{\lambda}(z) $ 
where function $ \mathcal{M}_{\lambda}(z) $ is 
defined in Eq.(\ref{mz}). To this end, we note that for a fixed $ z $, 
Eq.(\ref{mz}) can be viewed (up to a multiplier)
as the tail probability of a Poisson distribution with intensity $ \lambda M(z) $. Therefore, this 
function can be estimated using the large deviation result (\ref{Cramer_tail}), provided we substitute
$ \lambda \rightarrow \lambda M(z) $:
\beq
\label{Cramer_2}
\mathcal{M}_{\lambda}(z) =   e^{ \lambda T (M(z) - 1)} P \left( Po(\lambda T M(z)) > n_0 \right) \sim 
e^{ \lambda T (M(z) - 1)} e^{ - n_0 \left( \lambda T M(z) - \log \lambda T  - \log M(z) - 1 \right) }
\eeq 
We see that $ \mathcal{M}_{\lambda}(z) $, and hence its imaginary part, is exponentially small
in the limit $ s \rightarrow \infty $. On the other hand, the term $ \sim  \Delta \mathrm{Im} \,\, M_{\lambda} (x)$
in Eq.(\ref{main_c}) is only suppressed as a power of $ 1/s $, as will be clear below. Omitting exponentially suppressed contributions in Eq.(\ref{main_c}) and 
introducing a new variable $ y = x_0 x $ as before, we finally obtain 
\beq
\label{main_2}
\bar{F}_{\lambda}(s) = 
\frac{1}{\pi} \int_{0}^{\infty} \frac{dy}{y} e^{ - \hat{s} y + \lambda T \Psi(y/x_0)} 
\sin \left[ \frac{ \pi \omega \lambda T}{\Gamma(\alpha-1)} 
y^{\alpha-1} \right] 
\eeq
Eq.(\ref{main_2}) constitutes our second main result. We have derived it as an application of 
our general relation (\ref{main}). Up to exponentially suppressed terms, Eq.(\ref{main_2}) 
provides an exact expression for the large-$s$ behavior of a random sum of {\it i.i.d.} random
variables with power-law tails. As the integral (\ref{main_2}) converges rapidly for 
$ s \rightarrow \infty $, it can be evaluated very efficiently using quadratures. Alternatively,
it can be expanded into an asymptotic series, as we discuss next.

\subsection{Asymptotic expansion of the tail probability}

As was mentioned above, as the integral (\ref{main_2}) converges fast, a direct numerical 
integration is the most straightforward way to evaluate it.
However, to better understand the asymptotic behavior of the tail probability of a compound Poisson
loss process, it is useful to analyze an asymptotic expansion of the tail probability 
in the limit $ s \rightarrow \infty $ that stems from Eq.(\ref{main_2}). In particular, such asymptotic 
expansion allows one to analytically study the impact of a specification of the low-loss part of the 
loss distribution on the resulting VAR figures.

We start with an observation that for large $ s \rightarrow \infty $, the integral (\ref{main_2}) is 
dominated by small values of $ y $. Therefore, we can evaluate (\ref{main_2}) using Taylor 
expansions of functions $ \Psi(\cdot) $ and $ \sin(\cdot) $ that enter this expression. Using 
Eq.(\ref{R}), we have 
\beq
\label{Psi_2}
\Psi(y/x_0) = \left( 1 - \omega \right) M_1(-y/x_0) + \omega \Phi(1-\alpha, 2- \alpha, y) + 
\omega \Gamma(2-\alpha) \cos ( \pi \alpha) y^{\alpha-1}  - 1
\eeq 
Using the Taylor expansion of Kummer's function
\beq
\label{Taylor_Kummer}
\Phi(a,b,z) = 1 + \frac{a}{b} z + \frac{1}{2!} \frac{a(a+1)}{b(b+1)} z^2 + \ldots = \sum_{n=0}^{\infty}
\frac{1}{n!} \frac{a^{(n)}}{b^{(n)}} z^n 
\eeq
(where $ a^{(0)} = 1 $, $ a^{(n)} = a(a+1) \ldots (a+n-1) $) and a general Taylor expansion 
of the low-loss MGF $ M_1(-y/x_0) $
\beq
\label{Taylor_M1}
M_1( - y/x_0) = 1 + m_1 y + m_2 y^2 + \ldots
\eeq
we obtain the following small-$y$ expansions for functions entering Eq.(\ref{main_2}):
\bea
\label{Taylor_for_main}
&& \Psi(y/x_0) = \omega \Gamma(2-\alpha) \cos ( \pi \alpha) y^{\alpha-1} + 
c_1 y + 
c_2  y^2 + \ldots
\nonumber \\
&&\sin \left[ \frac{ \pi \omega \lambda T}{\Gamma(\alpha-1)} 
y^{\alpha-1} \right] = \frac{ \pi \omega \lambda T}{\Gamma(\alpha-1)} 
y^{\alpha-1} - \frac{1}{3!} \left( \frac{ \pi \omega \lambda T}{\Gamma(\alpha-1)} 
y^{\alpha-1} \right)^3 + \ldots
\eea
where 
\beq
\label{c12}
c_1 = \frac{\omega}{1!} \frac{1-\alpha}{2-\alpha} + (1-\omega) m_1\, , \; \; 
c_2 =  \frac{\omega}{2!} \frac{1-\alpha}{3-\alpha} + (1-\omega) m_2 
\eeq
As $ \Psi(y/x_0) \rightarrow 0 $ as $ y \rightarrow 0 $, we can additionally expand the exponent
in Eq.(\ref{main_2}) as $ \exp \left(\lambda T \Psi(y/x_0) \right) = 1 + \lambda T \Psi(y/x_0) + \ldots $.
Using this along with Eqs.(\ref{Taylor_for_main}), we obtain the following asymptotic expansion 
of the tail probability (\ref{main_2}):
\bea
\label{asympt_comp}
\bar{F}_{\lambda}(s) &=& \omega \lambda T \left( \frac{s}{x_0} \right)^{-(\alpha-1)} +
a_1 \left( \frac{s}{x_0} \right)^{-2(\alpha-1)}  +
a_2 \left( \frac{s}{x_0} \right)^{-\alpha}  
\nonumber \\  
&+& a_3 \left( \frac{s}{x_0} \right)^{-3(\alpha-1)} +    
a_4 \left( \frac{s}{x_0} \right)^{-(2\alpha-1)} + 
a_5 \left( \frac{s}{x_0} \right)^{-(\alpha+1)} + \ldots 
\eea
where 
\bea
\label{coeffs}
a_1 &=& \left( \omega \lambda T \right)^2 \frac{\Gamma(2-\alpha) \Gamma(2 \alpha -2)}{
\Gamma(\alpha-1)} \cos ( \pi \alpha)  = - \frac{1}{2} \left( \omega \lambda T \right)^2 
\frac{\left(\Gamma(2-\alpha) \right)^2}{
\Gamma(3 - 2 \alpha)} \nonumber \\
a_2 &=& \omega (\lambda T)^2 c_1 (\alpha-1) = \omega (\lambda T)^2 (\alpha -1) 
\left( \omega \frac{1-\alpha}{2-\alpha} + (1-\omega) m_1 \right) \nonumber \\ 
a_3 &=& - \frac{\pi^2}{3!} \Gamma(3 \alpha-3) \left( \frac{ \omega \lambda T}{\Gamma(\alpha-1)}
\right)^3 \\ 
a_4 &=& \left( \omega \lambda T \right)^2 \lambda T c_1 \frac{\Gamma(2-\alpha) \Gamma(2 \alpha -1)}{
\Gamma(\alpha-1)} \cos ( \pi \alpha) = 
\frac{1}{2} \left( \omega \lambda T \right)^2 \lambda T \frac{\left(\Gamma(2-\alpha) \right)^2}{
\Gamma(2 - 2 \alpha)} \left( \omega \frac{1-\alpha}{2-\alpha} + (1-\omega) m_1 \right)  
\nonumber  \\
a_5 &=& \omega (\lambda T)^2 \left( c_2 + \frac{\lambda T}{2} c_1^2 \right) \frac{\Gamma(\alpha+1)}{
\Gamma(\alpha -1)}  = \omega (\lambda T)^2 \alpha (\alpha-1) 
\left( c_2 + \frac{\lambda T}{2} c_1^2 \right)\nonumber 
\eea  
Note that if we only keep the leading term in Eq.(\ref{asympt_comp}), we obtain
\beq
\label{asympt_leading}
\bar{F}_{\lambda}(s) = \omega \lambda T \left( \frac{s}{x_0} \right)^{-(\alpha-1)} = 
\lambda T \bar{F}(s) \, , \; \; s \rightarrow \infty 
\eeq
which is a well-known result in the literature \cite{BK}. Eq.(\ref{asympt_leading}) is 
remarkable in that it shows 
that in the limit $ s \rightarrow \infty $, the tail of the compound distribution with a power-law tail
{\it decouples} from the low-loss region, as the parameter $ \omega \lambda T $ entering this expression is the Poisson frequency of large losses and is therefore not sensitive to any additional small loss 
events that happen after the model is initially calibrated.  

Now consider the structure of corrections $ \sim a_i $ in Eq.(\ref{asympt_comp}). First, note that 
when $ 1 < \alpha < 2 $, the term $ \sim a_1 $ is a leading correction, while for $ \alpha > 2 $, 
the term $ \sim a_2 $ is a leading correction. If $ 1 < \alpha < 2 $, i.e. the single loss distribution has 
an infinite mean, we see that the leading correction is still decoupled from the body of the 
distribution as it only depends on the combination $ \omega \lambda T $. On the other hand, 
if $ \alpha > 2 $ (so that the distribution has a finite mean), then the leading correction depends 
on the body of the distribution through its mean $ m_1 $. The sub-leading correction depends 
on $ m_1 $ if $ 1 < \alpha < 2 $, or on both mean and variance $ m_1 $ and $ m_2 $, if $ \alpha > 2 $.
If we set $ \omega = 1 $ (so that there is no low-loss "body" of the distribution), our expression for 
$ a_1 $ coincides with an expression given in \cite{Sahay}.  
 
\section{Tail probability of a random lognormal sum}
\label{Sect:Lognormal}

In this section, we analyze random sums of lognormal random variables using a similar approach 
to that developed in the previous section. The added difficulty of analysis of a lognormal distribution
is that its MGF is not available in an analytical form, but should rather be analyzed 
using its integral representation along with a saddle point approximation. As before, we 
start with the analysis of the tail probability of a single
loss distribution.

\subsection{MGF of a lognormal distribution in the left semi-plane}

A random variable $ Y = \exp(X) $ follows a lognormal distribution with mean $ \mu $ and variance
$ \sigma^2 $ if $ X \sim N (\mu, \sigma^2) $ is a normal random variable with the same mean and 
variance. The tail probability of a lognormal variable can be computed using elementary means:
\beq
\label{tail_logn_1}
\bar{F}(s) = P \left( Y > s \right) = P \left( X > \log s \right) = N \left( - \frac{ \log s - \mu}{\sigma} \right)
\eeq
where $ N(x) $ stands for the cumulative normal distribution. 

Now we represent the  tail probability (\ref{tail_logn_1}) as an inverse Laplace transform of its Laplace
transform:
\beq
\label{inv_Lap_logn}
\bar{F}(s) = \frac{1}{2 \pi i} \int_{-\varepsilon - i \infty}^{-\varepsilon + i \infty} dz e^{ sz} 
\bar{ \mathcal{F}}(z)
\eeq
where $ \bar{ \mathcal{F}}(z) $ stands for the Laplace transform of the tail probability
\beq
\label{Lap_logn}
\bar{ \mathcal{F}}(z) = \int_{0}^{\infty} dx e^{-zx}  N \left( - \frac{ \log x - \mu}{\sigma} \right) = 
- \frac{1}{z} \frac{1}{\sqrt{2 \pi \sigma^2}} \int_{0}^{\infty} \frac{dx}{x} \exp \left( - zx - 
\frac{1}{2 \sigma^2} \left( \log x - \mu \right)^2 \right)
\eeq
where we used integration by parts to get the second equation.  
Comparing Eq.(\ref{inv_Lap_logn}) with 
Eq.(\ref{tail_3}) and using (\ref{Lap_logn}), we find the MGF
\beq
\label{MGF_logn}
M(z) = \frac{1}{\sqrt{2 \pi \sigma^2}} \int_{0}^{\infty} \frac{dx}{x} \exp \left( - zx - 
\frac{1}{2 \sigma^2} \left( \log x - \mu \right)^2 \right)
\eeq
which of course can also be obtained directly from the definition of the MGF. 
Note that Eq.(\ref{MGF_logn}) defines the MGF $ M(z) $ for $ Re \, z \geq 0 $. 
Its value in the left semi-plane $ Re \, z < 0 $ is defined by 
analytical continuation as described below. 
Prior to doing this, we note that using the identity
\beq
\label{Laplace_theta}
\theta(x > s) = - \frac{1}{2 \pi i} \int_{-\varepsilon - i \infty}^{-\varepsilon + i \infty} 
\frac{dz}{z} e^{ z(s-x)}
\eeq
we can verify that substituting Eq.(\ref{Lap_logn}) into Eq.(\ref{inv_Lap_logn}), interchanging the 
orders of integrals over $ z $ and $ x $, and closing the contour in the right semi-plane, we 
reproduce (\ref{tail_logn_1}). 

Now instead of closing the integration contour in  
Eq.(\ref{inv_Lap_logn}) in the right semi-plane, we want to use analytical properties of 
the MGF (\ref{MGF_logn}) to produce a different expression for the tail probability 
(\ref{inv_Lap_logn}).
To this end, we note that the MGF (\ref{MGF_logn}) is an analytic function of $ z $ in a cut plane with
a branch cut for $ z \in [- \infty, 0] $. The branch cut singularity in the $ z $-plane arises due to 
a branch cut singularity of the integrand of (\ref{MGF_logn}) in the $ x $-plane due to 
multivaluedness of  
the logarithm. 

To find the MGF $ M(z) $ in the left semi-plane $ Re \, z < 0 $, one should  
analytically continue the integral (\ref{MGF_logn}). Let $ z = \xi e^{ i \theta} $ where
$ \xi = | z | \geq 0$. For $ \frac{\pi}{2}  < \theta < \frac{3 \pi}{2} $, the real part of $ z$ becomes negative, and 
to keep the integral convergent, we have to rotate the integration line in the $ x $-plane by 
$ - \theta $. This gives 
\beq
\label{MGF_logn_c}
M \left( \xi e^{ i \theta} \right) =  \frac{1}{\sqrt{2 \pi \sigma^2}} \int_{C} 
\frac{dx}{x} \exp \left( - x  \xi e^{i \theta} - 
\frac{1}{2 \sigma^2} \left( \log x - \mu \right)^2 \right)
\eeq
where the integration contour is a ray $\{ C: \, \arg(x) = - \theta, \, Re \, x \geq 0  \}$. Changing the variable to $ y = x e^{i \theta} $, we obtain 
\beq
\label{MGF_logn_c_2}
M \left( \xi e^{ i \theta} \right) =  \frac{1}{\sqrt{2 \pi \sigma^2}} \int_{0}^{\infty} 
\frac{dy}{y} \exp \left( - y  \xi - 
\frac{1}{2 \sigma^2} \left( \log y - \mu - i \theta \right)^2 \right)
\eeq
Note that $ M \left( \xi e^{ \theta} \right) \rightarrow 0 $ as $ \xi \rightarrow \infty $, 
therefore if we close the contour in Eq.(\ref{inv_Lap_logn}) 
in the left semi-plane, the integral over an infinite 
semi-circle vanishes by Jordan's lemma.  Deforming the integration contour 
in (\ref{inv_Lap_logn}) in a similar way to steps taken 
above in the derivation of Eq.(\ref{main}), 
the tail probability
is now expressed in terms of a discontinuity of $ \mathrm{Im} \,\, M(z) $ across the branch cut on $ z \in [-\infty,
0] $, as in Eq.(\ref{main}):
\beq
\label{main_logn}
\bar{F}(s) = - \frac{1}{2 \pi } \int_{0}^{\infty} dx \frac{ e^{- s x}}{x} \Delta  \mathrm{Im} \,\, M(x) 
\eeq 
The discontinuity across the cut 
can be found using Eq.(\ref{MGF_logn_c_2}):
\bea
\label{d_Im_MGF_logn} 
\Delta  \mathrm{Im} \,\, M(x) &=&
\mathrm{Im} \,\, M \left( x e^{ i \pi} \right) -  \mathrm{Im} \,\, M \left( x e^{ - i \pi} \right) \nonumber \\
&=&   2\frac{ \exp( \frac{\pi^2}{2 \sigma^2}) }{\sqrt{2 \pi \sigma^2}} \int_{0}^{\infty} 
\frac{dy}{y} \exp \left( - x y  - 
\frac{1}{2 \sigma^2} \left( \log y - \mu \right)^2  \right)
\sin \left( \frac{\pi}{\sigma^2} \left( \log y - \mu \right) \right) \nonumber \\
&=&   2\frac{\exp( \frac{\pi^2}{2 \sigma^2})}{\sqrt{2 \pi \sigma^2}} \int_{-\infty}^{\infty} 
dt  \exp \left( - x e^{\mu + t}  - 
\frac{1}{2 \sigma^2} t^2  \right)
\sin \left( \frac{\pi}{\sigma^2} t \right)  
\eea  
Note that $ \Delta  \mathrm{Im} \,\, M(x) \rightarrow 0 $ when $ x \rightarrow 0 $ (as 
the integrand in
(\ref{d_Im_MGF_logn}) becomes an odd function in this limit). Moreover, all terms of the Taylor 
expansion in the integrand result in converging integrals which all equal  
zero. This is a manifestation of 
non-analyticity of $M(z)$ at $z=0$.

To produce non-vanishing results for both $ \mathrm{Im} \, M \left(x e^{i \pi} \right)$ and
$ \mathrm{Re} \, M \left(x e^{i \pi} \right)$, 
we return to Eq.(\ref{MGF_logn_c_2}) where we now set $ \xi = x $, $ \theta = \pi $ and 
change the variable to $ z = \log \left( y e^{- \mu} \right) $:
\beq
\label{MGF_logn_c_3}
M \left( x e^{ i \pi} \right) = 
\int_{-\infty}^{\infty} 
\frac{dz}{\sqrt{2 \pi \sigma^2} } \exp \left[ \frac{1}{\sigma^2} 
\left( - \kappa e^{z}  - \frac{1}{2} (z-i\pi)^2  \right) \right] 
\equiv \int_{-\infty}^{\infty} 
\frac{dz}{\sqrt{2 \pi \sigma^2}}  \exp \left[ \frac{1}{\sigma^2} g(z) \right]
\eeq
where $ \kappa =  x \sigma^2 e^{\mu} $ and function $ g(z) $ is defined as follows:
\beq
\label{g_fun}
g(z) =  - \kappa e^{z}  - 
\frac{1}{2} (z-i\pi)^2 
\eeq
Stationary points of functions $ g(z) $ are zeros of the derivative
\beq
\label{dg}
g'(z) = - \kappa e^{z} - z + i \pi 
\eeq
Complex-valued stationary points can therefore be computed as 
\beq
\label{z_0}
z_0 = w + i \pi
\eeq
where $ w $ stands for real-valued solutions of the equation
\beq
\label{w_0_eq}
w = \kappa e^{w}
\eeq
In what follows, we restrict our analysis to the case when $ \kappa $ is bounded from above
as follows: 
\beq
\label{constr_kappa}
0 \leq \kappa \leq \frac{1}{e} \; \; \Leftrightarrow \; \;  0 \leq x \leq \frac{1}{\sigma^2} e^{- \mu - 1} 
\eeq
which seems sufficient for the analysis of asymptotic behavior at $ s \rightarrow \infty $, 
as the integral (\ref{main_logn}) is dominated by small values of $ x $ in this limit.
Provided (\ref{constr_kappa}) is satisfied, Eq.(\ref{w_0_eq}) has two 
real roots $ w_1 $ and $ w_2 $ that are expressed in terms of Lambert functions 
$ W_0(z) $ and $ W_{-1}(z) $ (see \cite{Corless}):
\bea
\label{w_01}
w_1 &=& - W_0(- \kappa) \, , \; \; -1 \leq - \kappa < \infty \, , \; \; W_0(- \kappa) \geq -1 
\nonumber \\
w_2 &=& - W_{-1}(- \kappa) \, , \; \; - \frac{1}{e} \leq - \kappa < 0  \, , \; \; W_{-1}(- \kappa) 
\leq -1 
\eea
We note that the first saddle point $ w_1 $ also appears in a saddle point analysis of $ M(z) $ in the 
right semi-plane $  \mathrm{Re}  \,  z \geq 0 $ (see Ref.\cite{Tella}), while in our approach we
concentrate on the behavior of $ M(z) $ in the left semi-plane  
$  \mathrm{Re}  \,  z \leq 0 $, where the second saddle point $ w_2 $ appears in addition to 
$ w_1 $. As will be seen shortly, it is the second saddle point $ w_2 $ that determines 
the imaginary part of $ M(z) $ along the branch cut at $ z \in [-\infty, 0 ] $.  

The Lambert functions $ W_0(x) $ and $ W_{-1}(x) $ have the following expansions for small
 $ x $ \cite{Corless}:
 \bea
 \label{Taylor_W_01}
 W_0(x) &=& \sum_{n=1}^{\infty} \frac{ (-n)^{n-1}}{n!} x^n = x - x^2 + \frac{3}{2} x^3 + \ldots
 \nonumber \\
 W_{-1}(x) &=& \log(-x) + \log \left( - \log( - x) \right) + \ldots
 \eea
Note that when $ \kappa \rightarrow 0 $ (i.e. $ x \rightarrow 0 $), we have $ w_1 \ll 1 $ 
and $ w_2 \gg 1 $, so that the two saddle points are well separated
from each other in this limit. 
Now compute the second derivative 
\beq
\label{g''}
g''(z_0) = - \kappa e^{z_0} - 1 = \kappa e^{w} - 1 = w - 1
\eeq
We see that $ g'' \left(w_1 \right) < 0 $, while $ g'' \left(w_2 \right) \gg 1 $. Therefore,
a saddle-point contour should run parallel to the real axis when passing through $ w_1 $, 
and parallel to the imaginary axis when passing through $ w_2 $ (see below for more detail). 
Note that because the two saddle points are well-separated in the limit $ \kappa \rightarrow 0 $,
calculations of the integral (\ref{MGF_logn_c_3}) amounts, in the saddle point approximation, to 
a sum of two separate saddle point integrals computed using  
quadratic approximations of $ g(z) $  
in the vicinity of points $ w_1 + i \pi  $ and $ w_2 + i \pi  $, respectively.
Furthermore, as $ g(w + i \pi) = w - \frac{1}{2} w^2 $, we see that  $ g(w_1 + i \pi) = O(1) $, 
while $ g(w_2 + i \pi) \rightarrow - \infty $ as $ w_2 \rightarrow \infty $ (i.e. $ \kappa \rightarrow 0$).
 Therefore a contribution of the second saddle point  $ w_2 $ to the real part  
$ \mathrm{Re} \, M \left(x e^{i \pi} \right)$ is exponentially suppressed in comparison to a 
contribution of the first saddle point $ w_1 $, and thus can be safely neglected in the limit 
$ \kappa \rightarrow 0 $\footnote{An exponentially suppressed contribution of the second
saddle point $ w_2 $ to the real part $ \mathrm{Re} \, M \left(x e^{i \pi} \right)$ arises due to 
integration over the right edge of the first arc in the steepest descent contour (\ref{C}), see 
below. }.

\begin{figure}[ht]
\begin{center}
\includegraphics[width=82.68mm,height=35.88mm]{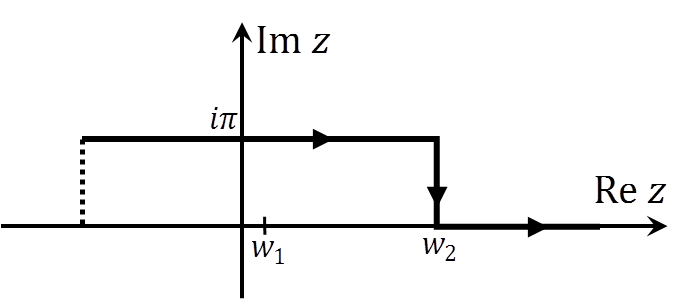}
\caption{Integration contour for integral (\ref{MGF_logn_c_3}).} 
\label{fig:zigzag}
\end{center}
\end{figure}

The saddle-point approximation is obtained using the standard arguments, see e.g. \cite{MW}. We expand 
$ g(z) $ in a Taylor series around the saddle point $ z_0 $ 
\beq
\label{T_1}
g(z) = g(z_0) +  \frac{1}{2!} g''(z_0) \left( z - z_0 \right)^2 + 
+ \frac{1}{3!} g'''(z_0) \left( z - z_0 \right)^3 + 
\frac{1}{4!} g^{(4)}(z_0) \left( z - z_0 \right)^4 + \ldots
\eeq
(where the first order term is omitted as $ g'(z_0) = 0 $), and introduce the following notation
\bea
\label{saddle_point_vars}
g''(z_0) &=& \rho e^{ i \theta} \;  \; \; ( \rho \geq 0 ) \nonumber \\
z - z_0 &=& t e^{ i \phi} \\
g(z) &=& u(x,y) + i v(x,y) \nonumber 
\eea
Note that $ ( \rho, \theta) = (1-w_1, \pi ) $ for $ z_0 = w_1 + i \pi $, 
and $ (\rho, \theta) = ( w_2 - 1, 0) $ for 
$ z_0 = w_2 + i \pi $. All higher-order derivatives $ g''', g^{(4)}, \ldots $ are real-valued at both 
saddle points. Using this in Eq.(\ref{T_1}), we obtain
\bea
\label{g_z_expansion}
u(x,y) &=& u(x_0,y_0) + \frac{1}{2} \rho t^2 \cos \left( \theta + 2 \phi \right) + 
\sum_{n=4,6 \ldots}\frac{1}{n!} g^{(n)} \left( z_0 \right) t^n \cos \left( n \phi \right) 
\nonumber \\
v(x,y) &=& v(x_0,y_0) +  \frac{1}{2} \rho t^2 \sin \left( \theta + 2 \phi \right) +
\sum_{n=3,5 \ldots}\frac{1}{n!} 
g^{(n)} \left( z_0 \right) t^n \sin \left( n \phi \right) 
\eea
A steepest descent path is therefore 
defined by the constraint $ \cos \left( \theta + 2 \phi \right)  =  -1 $.
This produces $ \phi = 0 $ for $ w = w_1 $, and 
$ \phi = \pi/ 2 $ for $ w = w_2 $. 
As $ dz = e^{i \phi} d t $ near a saddle point, this means that $ dz = t $ near $ w_1$, 
and $ dz = i dt $ near $ w_2$.
Therefore, the steepest descent contour should 
run parallel to the {\it real} axis at $ z = w_1 +  i \pi $, and parallel to 
the {\it imaginary} axis at 
$ z =  w_2 +  i \pi $, as was already mentioned above. 
A saddle point contour $ C $ that passes through both saddle points and works for both
the imaginary and real parts of the integral (\ref{MGF_logn_c_3}) can be obtained 
as a composition of three straight lines (see Fig.~\ref{fig:zigzag}):
\bea
\label{C}
C \, = \, \left\{ \begin{array}{clcr}
x + i \pi,  &  \mbox{if $ - \infty <  \mathrm{Re}  \,  z <  w_2 $}  \\ 
w_2 + i \theta, \; 
&  \mbox{if $ \mathrm{Re}  \,  z = w_2 \, , \;  0 \leq \theta \leq \pi $} \\
x,  &  \mbox{if $ \mathrm{Re}  \,  z > w_2 $} 
\end{array} \right. 
\eea 
Note that as 
$ \cos \left( \theta + 2 \phi \right)  =  -1 
 $ on the
 saddle point contour
$ C $, the second equation in (\ref{g_z_expansion}) implies
that $ v(x,y) = v(x_0, y_0) $ along this contour. In our case, $ v(x_0, y_0)  = 0 $ for both 
saddle points, therefore $ v(x, y) =  \mathrm{Im}  \, g(z) = 0 $ on $ C $.

The last observation that $  \mathrm{Im}  \, g(z) = 0 $ along the 
steepest descent path $ C $ implies that the contribution of the saddle point $ w_1 + i 
\pi $ (more precisely, the whole contribution of the first arc in (\ref{C})) 
drops off in the calculation of the imaginary part, because both the integrand and the integration 
contour become real for this integral after a change of variable $ z = x + i \pi $:
\[
\mathrm{Im} \,  \int_{-\infty + i \pi}^{w_2 + i \pi} 
\frac{dz}{\sqrt{2 \pi \sigma^2} } \exp \left[ - \frac{ \kappa e^{z}  + (z-i\pi)^2 /2}{\sigma^2} 
\right] 
= \mathrm{Im} \,  \int_{-\infty}^{w_2} 
\frac{dx}{\sqrt{2 \pi \sigma^2} } \exp \left[ - \frac{ - \kappa e^{x}  + \frac{1}{2} x^2 }{\sigma^2} 
\right] = 0
\]
Therefore, even though the contribution of the second saddle 
point $ w_2 + i \pi $ is exponentially suppressed when computing the {\it real} part 
$ \mathrm{Re} \, M \left(x e^{i \pi} \right) $ (see below),
as the saddle point contour is 
complex-valued in the vicinity of $ w_0^{(2)} + i \pi $,
the second saddle point $ w_2 $ is the only 
one that determines 
the {\it imaginary} part of $ M(z) $ along the branch cut. The dominant contribution to the 
imaginary part  $ \mathrm{Im} \, M \left(x e^{ i \pi} \right) $ comes from integration over the 
second arc in the contour (\ref{C}) (the integral over the third arc is exponentially suppressed 
in comparison to a contribution of the second arc, see 
below). Note that the contribution of $ w_1 $ to the imaginary part of $ M(z) $ cancels out precisely
due to a judicious choice of the integration contour (\ref{C}). Unless such cancellation occurs, an
accurate calculation of $ \mathrm{Im} \, M \left(x e^{ i \pi} \right)$ would be impossible for 
all practical 
purposes as it would require computing an exponentially suppressed quantity
with a large "numerical noise" that would be driven by an error 
in calculation of the contribution of the first arc in (\ref{C})\footnote{Our 
approach was inspired by contour integration methods that have been used for
 similar problems of estimating exponentially suppressed integrals in 
quantum mechanics (see Ref.\cite{Landau}, \$ 51) and in quantum field theory
\cite{ZJ}.}. The saddle point approximation 
to this integral produces the following result:    
\beq
\label{Im_M}
\mathrm{Im} \, M \left(x e^{i \pi} \right) = 
- \frac{1}{2} \frac{ \exp \left( \frac{1}{\sigma^2} \left( 
w_2 - \frac{1}{2} w_2^2 \right) \right) }{\sqrt{w_2 - 1}} 
\left[  1 + \frac{ \sigma^2}{8} \frac{ w_2}{ (w_2 - 1)^2} + O \left( w_2^{-2} \right)  \right]
\eeq   
and $ \mathrm{Im} \, M \left(x e^{- i \pi} \right) = - \mathrm{Im} \, M \left(x e^{i \pi} \right)  $. 
The extra factor $ \frac{1}{2} $ in Eq.(\ref{Im_M}) is due to the fact that the saddle point 
contour (\ref{C}) involves only half\footnote{After approximating the finite integration
limits $ [w_2- i \pi, w_2 + i \pi ] $ by infinite limits
$ [ w_2 - i \infty, w_2 + i \infty ] $, which is justified in the saddle
 point approximation in the 
limit $ w_2 \rightarrow \infty $. The difference between results obtained with an 
infinite and finite intervals is exponentially suppressed as $ \kappa \rightarrow 0 $,
and thus should be omitted as long as we omit other exponentially suppressed terms 
in our derivation.}  
of the line $ [ w_2 - i \infty, w_2 + i \infty ] $ 
which would otherwise be obtained as a saddle point contour for a problem with a 
single saddle 
point at $ w_2 $ and $ g'' \left(w_2 \right) > 0 $.  Note that Eq.(\ref{Im_M}) is 
an asymptotic expansion valid for small $ 1/w_2 
\rightarrow 0 $, i.e. for $ \kappa \rightarrow 0 $. 
Also note non-analyticity of this expression in $ w $ (and hence in $ \kappa $), which 
is due to the branch cut singularity of the square root function in (\ref{Im_M}).

To show that the contribution $ I_3 (\kappa) $ of the third arc of the saddle point 
contour (\ref{C}) to 
the integral (\ref{MGF_logn_c_3}) is negligible as was promised above, we 
use the following inequalities:
\[
|I_3(\kappa)|<
\int_{w_2}^{\infty}
\frac{dx}{\sqrt{2 \pi} \sigma}  \exp \left( - \kappa e^{x}  - 
\frac{1}{2 \sigma^2} (x^2-\pi^2)  \right)<
\frac{\sigma}{2\sqrt{2 \pi} w_2 }  \exp \left( - \frac{w_2}{\sigma^2}  - 
\frac{1}{2 \sigma^2} (w_2^2-\pi^2)  \right) 
\]
The second inequality here is obtained by noting that the maximum of the function
$ \phi(x) = - \kappa e^x - \frac{1}{2} \left( x^2 - \pi^2 \right) $ is 
attained at the left boundary $ x = w_2 $, and expanding $ \phi(x) $ to linear order 
$ \phi(x) = \phi(w_2) + \phi'(w_2) \left( x  - w_2 \right) + \ldots $ to evaluate the 
integral.  
We see that $\mathrm{Im} \, I_3(\kappa)$ is exponentially suppressed 
in comparison with the contribution calculated in 
Eq. (\ref{Im_M}), and thus can be neglected in the limit 
$w_2 \to\infty$ (i.e. $\kappa\to 0$). Likewise,  $\mathrm{Re} \, I_3(\kappa)$
is exponentially suppressed relatively to a contribution due to the first arc of contour 
(\ref{C}), as we will see next.  

Turning to the calculation of the real part of $ M(z) $ along the branch cut, it is 
now only
the first saddle point $ w_1 $ (i.e. the first arc of the contour (\ref{C})) 
that determines the real part $ 
\mathrm{Re} \, M \left(x e^{i \pi} \right) $, up to exponentially suppressed terms. 
The saddle point approximation produces the following result: 
\beq
\label{Re_M}
\mathrm{Re} \, M \left(x e^{i \pi} \right) = \frac{ \exp \left( \frac{1}{\sigma^2} \left( 
w_1 - \frac{1}{2} w_1^2 \right) \right) }{\sqrt{1 - w_1}} 
\left[  1 + \frac{ \sigma^2}{8} \frac{ w_1}{ (1 - w_1)^2} + O \left( w_1^{2} \right)  \right]
\eeq   
 Expressions (\ref{Im_M}) and (\ref{Re_M}) will be used below 
 to compute the tail distribution of a random sum of lognormal variables. Before doing that, 
 we want to verify that Eq.(\ref{Im_M}) reproduces the right asymptotic behavior of a tail of 
 a single lognormal loss when substituted in the general expression (\ref{main_logn}).

 \subsection{Tail probability of a single lognormal distribution}
 
 In this section, we check that our contour integral representation correctly reproduces the 
 asymptotic behavior of the tail of the lognormal distribution (\ref{tail_logn_1}):
 \beq
\label{tail_logn_11}
\bar{F}(s) = N \left( - \frac{ \log s - \mu}{\sigma} \right) = \frac{1}{\sqrt{2 \pi}} \frac{\sigma}{
\log s - \mu} \exp \left( - \frac{1}{2 \sigma^2} \left( \log s - \mu \right)^2 \right) + \ldots
\eeq
As was shown above, it is only the second point $ w_2 $ that determines the imaginary
 part $ \mathrm{Im} \, M(z) $ along the branch cut. Therefore, to lighten the notation, in this section
 we omit the index of the saddle point and write it simply as $ w = w_2 $ without any confusion.
 Note the integral (\ref{main_logn}) involves integration over $ x $, while the asymptotic expression
 (\ref{Im_M}) is valid for an arbitrary {\it fixed} (and small) value $ x \rightarrow 0 $. The integral 
 (\ref{main_logn}) could therefore be calculated, after a discretization on 
 a grid $ \{x_i \} $ ($ i = 1, 2, \ldots $), by computing the saddle 
 point $ z_0(x_i) $ for each value $ x_i $ on the grid.  However, it is much more 
 convenient to change the integration variable $ x \rightarrow w $ using Eq.(\ref{w_0_eq})
 as a definition of the change of variables, thus 
 avoiding a re-calculation of the saddle point for different values on the grid.
The $ x $ variable is found in terms of $ w  $ as follows:
\beq
\label{x_from_w}
x = \frac{1}{\sigma^2} e^{ - \mu} w \exp( - w) 
\eeq   
and the Jacobian $ \mathcal{J}  $ of the transformation is
 \beq
 \label{Jacobian}
 \mathcal{J} =  \left| \frac{\partial x}{ \partial w} \right| = \frac{1}{\sigma^2} e^{ - \mu - w} 
 (w-1) 
\eeq
Using this in Eq.(\ref{main_logn}) along with Eq.(\ref{Im_M}) and truncating the integral
at $ x = \frac{1}{\sigma^2} e^{ - \mu - 1} $ (which is justified in the limit $ s \rightarrow \infty $), 
we obtain
 \bea
 \label{main_logn_single}
\bar{F}(s) &=& -\frac{1}{2 \pi } \int_{0}^{\infty} dx \frac{ e^{- s x}}{x} \Delta  \mathrm{Im} \,\, M(x) 
\simeq -\frac{1}{2 \pi } \int_{0}^{\frac{1}{\sigma^2} e^{-\mu - 1}}
 dx \frac{ e^{- s x}}{x} \Delta  \mathrm{Im} \,\, M(x)
\nonumber \\
&=&  \frac{1}{2 \pi } \int_{1}^{\infty}  
\frac{dw}{w} \sqrt{w-1}
 \exp \left( \frac{1}{\sigma^2} \left( - s w e^{ - \mu - w} +  
w - \frac{1}{2} w^2 \right) \right)  
\left[  1 + \frac{ \sigma^2}{8} \frac{ w}{ (w - 1)^2} + \ldots \right]  \nonumber \\
&\equiv& \frac{1}{2 \pi } \int_{1}^{\infty} 
\frac{dw}{w} \sqrt{w-1}
 \exp \left( \frac{1}{\sigma^2} \Phi(w) \right)  
\left[  1 + \frac{ \sigma^2}{8} \frac{ w}{ (w - 1)^2} + \ldots \right]  
\eea
where 
\beq
\label{Phi_w}
\Phi(w) =  - s w e^{ - \mu - w} +  
w - \frac{1}{2} w^2 
\eeq
This integral can be computed in the limit 
$ s \rightarrow \infty $ using the saddle point approximation. 
Saddle points are found as solutions of the equation
\beq
\label{phi_der}
\frac{d  \Phi (w)}{dw}  = \left( w - 1 \right) \left( s e^{ - w - \mu} - 1 \right) = 0
 \eeq
This equation has two solutions
$w_1 = 1 $ and $ w_2 = \log s - \mu $, 
however, it is only the second solution $ w_2 $ that corresponds to the minimum of $ \Phi(w) $, as 
can be easily checked by differentiation of Eq.(\ref{phi_der}). Therefore, we pick $ w_2 $ as the saddle 
point 
\beq
\label{w_0}
w_0  = \log s - \mu
\eeq
Note that $ w_0 \gg 1 $ when $ s \rightarrow \infty $. We expand $ \Phi(w) $ around this point:
\beq
\label{Phi_expan}
\Phi(w) = \Phi(w_0) + \frac{1}{2} \Phi''(w_0) \left( w - w_0 \right)^2 + \ldots
\eeq
We have
\bea
\label{Phi_00}
\Phi(w_0) &=& - \frac{1}{2} \left( \log s - \mu \right)^2 \nonumber \\
\Phi''(w_0) &=& - \log s + \mu + 1 
\eea
Omitting correction terms in (\ref{main_logn_single}), the leading 
order saddle point approximation reads
\beq
\label{main_logn_single_2}
\bar{F}(s) = \frac{1}{\sqrt{2 \pi}} \frac{\sigma}{
\log s - \mu} \exp \left( - \frac{1}{2 \sigma^2} \left( \log s - \mu \right)^2 \right) + \ldots
\eeq
which coincides with Eq.(\ref{tail_logn_11}). We have therefore validated our 
result (\ref{Im_M}) using the known expression for the tail probability of a single-loss lognormal 
distribution. 

\subsection{Tail probability of a spliced distribution with a lognormal tail}

Now we would like to discuss a practically important case of of a spliced loss severity distribution 
with a low-loss "body" with a MGF $ M_1(z) $ and a lognormal tail that starts
at a junction point $ x_0 $. The MGF $ M_s(z) $ of the spliced distribution is therefore
\beq
\label{MGF_spliced_logn}
M_s (z) = \omega \tilde{M}(z) + (1-\omega) M_1(z)
\eeq
where $ \tilde{M}(z) $ stands for an MGF of a lognormal distribution truncated from below at $ x_0 $.
As before, we assume that the MGF $ M_1(z) $ of the "body" is an analytic function 
of $ z $, i.e. this distribution has all moments. The normalized truncated lognormal distribution has the following pdf:
\beq
\label{truncated_logn_pdf}
\tilde{p}(x) = \theta(x \geq x_0) \frac{\nu }{\sqrt{2 \pi \sigma^2}} \frac{1}{x}
\exp \left( - \frac{1}{2 \sigma^2} \left( \log x - \mu \right)^2  \right), \; \; 
\nu = \frac{1}{ N \left( - \frac{ \log x_0 - \mu}{\sigma} \right)} 
\eeq
The MGF of a truncated lognormal distribution reads
\beq
\label{MGF_trunc}
\tilde{M}(z) = \nu
\int_{x_0}^{\infty} \frac{dx}{\sqrt{2 \pi \sigma^2} x} 
\exp \left( - x z - \frac{1}{2 \sigma^2} \left( \log x - \mu \right)^2  \right)  
\eeq
Note that apart from the constant multiplier $ \nu $, the integral (\ref{MGF_trunc}) only differs in the 
lower integration limit ($x_0 $ instead of $ 0 $) from the integral (\ref{MGF_logn}) that 
defines the MGF of an un-truncated lognormal distribution. Repeating steps leading to 
Eq.(\ref{MGF_logn_c_3}), but this time with the low bound of $ x_0 $, we obtain
\beq
\label{MGF_logn_c_3_s}
\tilde{M} \left( x e^{ i \pi} \right) = 
\nu \int_{\bar{w}_0}^{\infty} 
\frac{dz}{\sqrt{2 \pi \sigma^2}}  \exp \left[ \frac{1}{\sigma^2} g(z) \right] \, , \; \; 
\bar{w}_0 = \log x_0 - \mu 
\eeq
where function $ g(z) $ was defined in Eq.(\ref{g_fun}). The latter integral can now be 
evaluated using the saddle point approximation as was done above for an un-truncated 
lognormal distribution. 

However, a detailed re-calculation is not necessary in this case. 
Recall that results of a saddle point approximation are not sensitive, 
to the leading order, to precise values of integration bounds as long as the latter 
are far away from a 
saddle point. This implies that  
as long as $ \bar{w}_0  \ll w_2 $ (where $ w_2 $ is defined in 
Eq.(\ref{w_01})), the result for the 
imaginary part of $ M_s \left(x e^{i \pi} \right) $ is the same (up to a constant multiplier) 
as for $ M\left(x e^{i \pi} \right) $ 
(see Eq.(\ref{Im_M})):
\beq
\label{Im_Ms}
\mathrm{Im} \, M_s \left(x e^{i \pi} \right) = -\frac{\omega \nu}{2} 
\frac{ \exp \left( \frac{1}{\sigma^2} \left( 
w_2 - \frac{1}{2} w_2^2 \right) \right) }{\sqrt{w_2 - 1}} 
\left[  1 + O \left( w_2^{-1} \right)  \right]
, \; \; \left( \bar{w}_0 \ll w_2 \right)
\eeq  
To compute the real part of $ M_s \left(x e^{i \pi} \right) $, we have to 
separately consider two cases: $ \bar{w}_0 < w_1  $ and $ \bar{w}_0 > w_1 $. In the first case, 
the first saddle point $ w_1 $ in Eqs.(\ref{w_01}) lies inside of 
the integration interval, and the result for 
$ \mathrm{Re} \, M_s \left(x e^{i \pi} \right) $ reads (see Eq.(\ref{Re_M}))
\beq
\label{Re_Ms_1}
\mathrm{Re} \, M_s \left(x e^{i \pi} \right) = 
(1-\omega) \mathrm{Re} \, M_1 \left(x e^{i \pi} \right) 
+ \omega \nu 
\frac{ \exp \left( \frac{1}{\sigma^2} \left( 
w_1 - \frac{1}{2} w_1^2 \right) \right) }{\sqrt{1 - w_1}} 
\left[  1 + O \left( w_1 \right)  \right]
, \; \; \left( \bar{w}_0 < w_1 \right)
\eeq     
On  the other hand, in the second case $ \bar{w}_0 > w_1 $, the saddle point 
$ w_1 $ lies outside of the integration interval. In this case, the maximum of the integrand is 
attained at the left boundary of the integration interval.  The asymptotic expression in this case
reads
\beq
\label{Re_Ms_2}
\mathrm{Re} \, M_s \left(x e^{i \pi} \right) = (1-\omega) \mathrm{Re} \, M_1 \left(x e^{i \pi} \right) 
+ \frac{\omega \nu \sigma}{\sqrt{2 \pi}} 
 \frac{ \exp \left( \frac{1}{\sigma^2} \left( 
\bar{w}_0 - \frac{1}{2} \bar{w}_0^2 \right) \right) }{\bar{w}_0 - \kappa e^{\bar{w}_0}  } 
\left[  1 + O \left( w_1 \right)  \right]
, \; \; \left( \bar{w}_0 > w_1 \right)
\eeq     
Note that while in the previous expression (\ref{Re_Ms_1})   
the second term depends on $ x $ through a dependence of   
$ w_1 $ on $ x $ (see Eqs.(\ref{w_01})), the second term 
in Eq.(\ref{Re_Ms_2}) is a constant in $ x $, so that the $ x$-dependence of 
$ \mathrm{Re} \, M_s \left(x e^{i \pi} \right) $ in the case  $ \bar{w}_0 > w_1 $
arises solely due to the $ x $-dependence of the MGF $ M_1 \left(x e^{i \pi} \right) $ of the 
"body" of the distribution.

\subsection{Tail probability of a compound distribution with a lognormal tail}

In this section, we use our general relation (\ref{main}) along with asymptotic 
relations (\ref{Im_Ms})-(\ref{Re_Ms_2}) in order to compute the asymptotic expansion of 
the tail probability of a compound distribution where a single loss distribution is given by 
a spliced distribution with a low-loss "body" with a MGF $ M_1(z) $ and a lognormal tail that starts
at a junction point $ x_0 $. 

The MGF of a compound distribution with a Poisson frequency distribution is therefore
\beq
\label{MGF_comp_logn}
M_{\lambda} (z) =  e^{ \lambda T ( M_s (z) - 1)}
\eeq 
where $ M_s(z) $ is defined in Eq.(\ref{MGF_spliced_logn}). 

Recall that Eq.(\ref{main}) can only be applied for MGFs that grow not 
faster than $ e^{ - x_0 z} $ in the left semi-plane. This is the case in the 
present setting as $ M(z) \rightarrow 0 $ when $ z = R e^{i \theta} $ with $ R \rightarrow 
\infty $ and $ \frac{\pi}{2} \leq \theta < \frac{3 \pi}{2} $ (see
Eq.(\ref{MGF_logn_c_2})), which 
means that $ M_{\lambda}(z) $ is bounded in the left semi-plane.  

The discontinuity of the imaginary 
part of $ M_{\lambda} (z) $ across the branch cut at $ z \in [ - \infty, 0] $ reads
\beq
\label{Delta_M_lam}
\Delta \, \mathrm{Im} \, M_{\lambda} \left(x e^{i \pi}\right) = 2 e^{ \lambda T \left( \mathrm{Re} 
M_s \left(x e^{i \pi} \right) - 1 \right) } \sin \left[ \lambda T \, 
\mathrm{Im} \, M_s \left(x e^{i \pi} \right) \right]
\eeq
Using this in Eq.(\ref{main}), we obtain
\beq
\label{main_compound}
\bar{F}_{\lambda}(s) = - \frac{1}{\pi } \int_{0}^{\infty} \frac{dx}{x} e^{ - sx  + \lambda T \left( \mathrm{Re} 
M_s \left(x e^{i \pi} \right) - 1 \right) } \sin \left[ \lambda T \, 
\mathrm{Im} \, M_s \left(x e^{i \pi} \right) \right] 
\eeq 
Eq.(\ref{main_compound}) together 
with Eqs.(\ref{Im_Ms})-(\ref{Re_Ms_2}) constitute our third main result 
that provides a rapidly convergent integral for a compound distribution of single loss 
distributions with lognormal tails. Similar to Eq.(\ref{main_2}), this integral is 
dominated by small values of $ x $ as $ s \rightarrow \infty$, therefore we are justified
in using asymptotic relations (\ref{Im_Ms})-(\ref{Re_Ms_2}) to numerically evaluate the 
integral (\ref{main_compound}) in this limit. Also similar to Eq.(\ref{main_2}), the
low-loss component $ f_1(x) $ effectively enters Eq.(\ref{main_compound}) only via 
its lowest moments, leading to a decoupling of the tail behavior of the compound 
distribution from individual small losses.

\section{Compound distributions of compound heavy-tailed distributions}

In this section, we briefly discuss possible extensions of the present formalism 
to compute tail probabilities of compound loss distributions where individual 
components are themselves compound distributions made of individual loss distributions
with heavy tails. Such problem arise, in particular, when one computes total VAR of
a financial institution having a number of different UoMs.

Let $ X_1, \ldots, X_N $ ($ i = 1, \ldots, N $) be total losses in  $ N $ different
UoMs. Let $ M_{\lambda}^{(i)}(z) $ be corresponding 
MGF's for compound loss distributions for these
UoMs, and $ \lambda_i $ be their Poisson frequencies. 
The simplest assumption about the joint distribution of losses in all UoMs is to assume
independence between them. Such assumption can be justified on the grounds of 
diffuculties of accurate estimation of correlation measures for operational losses, as 
well as the absence of any obvious mechanisms that would induce loss correlations 
between different UoMs. In this case, the MGF of the total loss 
$ X = X_1 + \ldots + X_N $ is given by the product of individual MGFs. Using 
Eq.(\ref{MGF_c}), we obtain  
\beq
\label{MGF_cc}
M_{\lambda} (z) = \prod_{i=1}^{N} M_{\lambda}^{(i)}(z) = 
 \prod_{i=1}^{N} \exp \left[ \lambda_i T \left( M_i(z) - 1 \right) \right]  
= \exp \left[ \sum_{i=1}^{N} \lambda_i T ( M_i(z) - 1) \right]
\eeq
As this expression has the same functional form as Eq.(\ref{MGF_c}), the 
tail probability of the compound compound distribution can be computed using the 
same relation (\ref{main}) as was used above to compute 
tail probabilities of individual compound distributions.

Alternatively, if explicit modeling of dependencies between individual compound
losses of different UoMs is deemed desired or necessary, there are multiple ways to 
introduces such dependences in a tractable way. One simple approach is to promote
the Poisson intensities $ \lambda_i $ into stochastic variables 
$ \lambda_i({\bf Z}) $ that depend on a low-dimensional set of $ M $ common stochastic factors 
$ \bf{Z} $. When one conditions on a 
realization of $ {\bf Z} $, individual compound losses become
independent, therefore the conditional tail probability 
$ \bar{F}_{\lambda} \left(s| {\bf Z} \right) $ will still be given by
Eq.(\ref{main}) for each fixed value of $ {\bf Z} $, and the unconditional tail 
probability would be obtained as follows:
\beq
\label{tail_cc}
 \bar{F}_{\lambda} \left(s| {\bf Z} \right) = \int d {\bf Z} p( {\bf Z}) 
 \bar{F}_{\lambda} \left(s| {\bf Z} \right)
\eeq
where $  p( {\bf Z}) $ is the probability density of $ {\bf Z} $.

Finally, some standard copula models such as e.g. the Gaussian- or Student t-copula
can be cast into an equivalent factor framework which ensures 
conditional independence of individual component losses. The conditional distribution
of cumulative loss can be computed analytically using the saddle point method 
\cite{IH_saddle} in combination with methods presented above to compute marginal 
compound distributions. Details of such construction will be presented elsewhere.

\section{Application to operational risk}

In this section, we consider applications of our approach to calculation of 
the Operational Value at Risk (VAR) for a financial institution. 
We focus on the case where the large-loss component 
of an individual loss distribution is given by a power law distribution which is 
characterized by two parameters $ \alpha $ and $ x_0 $.
Once the threshold value of $ x_0 $ is specified, the value of $ \alpha $ can be obtained 
using the expression obtained by the maximum likelihood 
method (see \cite{Clauset_2007}):
\beq
\label{alpha}
\hat{\alpha} = 1 + N \left[ \sum_{i=1}^{N} \log \frac{x_i}{x_{0}} \right]^{-1}
\eeq
and the standard error of $  \hat{\alpha} $ is 
\beq
\label{alpha_std}
\sigma_{\alpha} = \frac{\hat{\alpha} - 1}{\sqrt{N}} + O \left( \frac{1}{N} \right)
\eeq

We illustrate our method with synthetic data produced using realistic model 
parameters that 
are similar to what is often observed in calibration to real world datasets. 

We used the 
following values of parameters: $u=40,000$, $x_0=100,000$, $\alpha=2.2$, $\lambda=20$, $\omega=0.35$. The mean and variance of the body
distribution are   63,877.92 and  $2.625 \cdot 10^8$, which corresponds to rescaled moments $m_1=-0.6388$ and $m_2=0.0131$.

\begin{figure}[ht]
\begin{center}
\includegraphics[width=111.2mm,height=72.1mm]{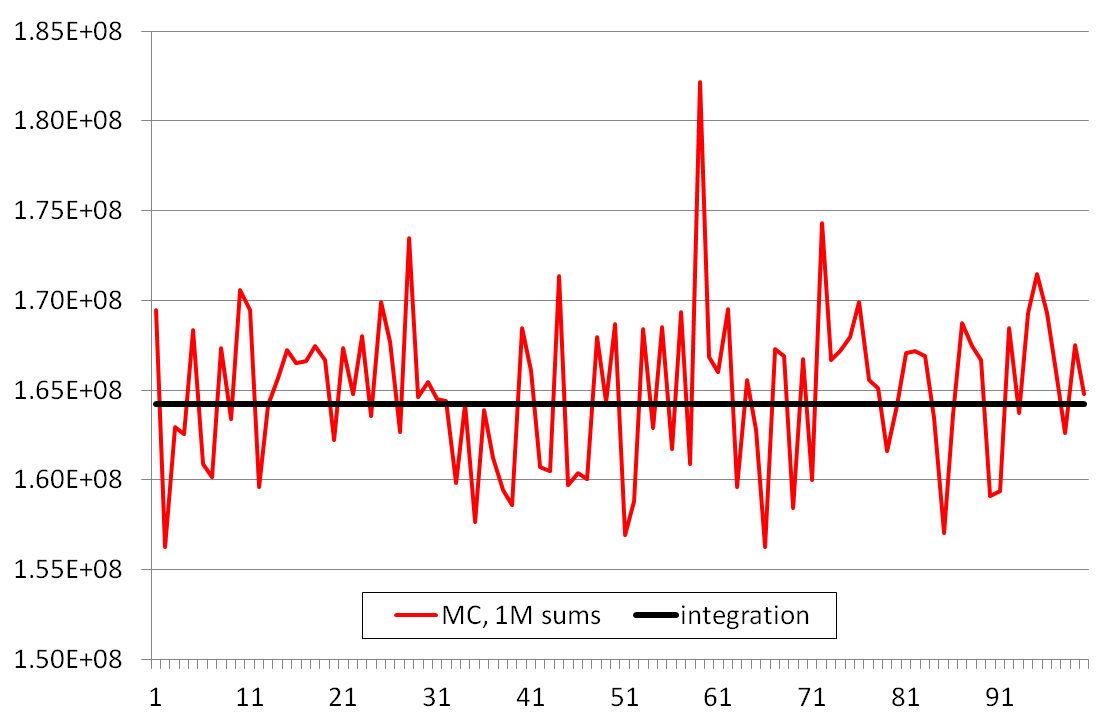}
\caption{Integration vs MC simulation.} 
\label{fig:MCvsIntegration}
\end{center}
\end{figure}

The result obtained with our approach for the percentile level of 99.9\% is shown in 
 Fig.~\ref{fig:MCvsIntegration} where we also show results of MC simulation with 100 runs, each including 1,000,000 aggregate losses. We note that
even for such a large number of MC trials the MC results are still quite noisy, which, in particular, makes computation of sensitivities difficult in the MC setting.
Our analytical approach, by construction, is free of such defficiencies and allows accurate computation of both quantiles and their sensitivities.

\section{Summary}

We have presented 
an analytical approach to computation of tail probabilities of compound 
heavy-tailed distributions, which 
is based on the contour integration method, and gives rise to a representation of the 
tail probability of a compound distribution in the form of a rapidly convergent real-valued one-dimensional integral. The latter integral 
can be evaluated in quadratures, or alternatively represented as an asymptotic
expansion. While we only considered the case of a compound Poisson distribution where individual components have power-law or lognormal tails, our method can be extended to other settings with different specifications of individual component and/or frequency distribution, as long as its moment
generating function has a branch cut singularity in the complex plane. We believe that the method proposed in this paper can offer a viable alternative to "brute-force" numerical methods such as Monte Carlo, FFT or the Panjer recursion. Interestingly, this alternative appears  
especially attractive for high percentile levels, where these 
traditional approaches struggle, while the contour integration method starts to shine even more (in the sense that a convergent integral defining the tail distribution converges even faster as the percentile level increases). 

\section*{Acknowledgements}

I would like to thank Sergey Malinin for numerous discussions and collaboration on an earlier version of this manuscript, including in particular sharing his ideas for calculations presented in Sect.~\ref{Sect:Lognormal}. I thank Andrey Itkin for helpful comments.   


\def\thesection{A}	
\setcounter{equation}{0}
\def\theequation{\thesection.\arabic{equation}}

\section*{Appendix A: Large deviations and Poisson tail probability}

To apply the large deviation theory (see e.g. \cite{Varadhan}) to estimation of the Poisson tail probability
\beq
\label{Poisson_tail}
P \left(Po(\lambda T) > n \right) = \sum_{k=n+1}^{\infty} \frac{(\lambda T)^k}{k!} e^{- \lambda T}
\eeq
we use the equivalence of the Poisson event $ Po(\lambda T) = n $ to  
the event of having $ n $ arrivals by time $ t = 1 $. If $ Z_i $ with $ i = 1, 2, \ldots $ stand for exponentially distributed 
random interarrival times, then $ P \left( Po(\lambda T) > n \right) $ is the same as the 
probability of having the total arrival time for $ n $ events to be less less than $ t = 1 $
\beq
\label{Poisson_Z}
P \left( Po(\lambda T) > n\right) = 
P \left( Z_1 + \ldots + Z_{n} < 1 \right) =  
P \left( \frac{Z_1 + \ldots + Z_{n}}{n} < \frac{1}{n}
 \right)
\eeq
To estimate this probability, we apply Cramer's theorem \cite{Varadhan}. We compute 
the rate function
\beq
\label{rate_fun}
I(z) = \sup_{x > 0} \left( z x - \log \mathbb{E} e^{x Z_1} \right) = \sup_{x > 0} \left( z x - \log \frac{\lambda T}{
\lambda T - x} \right) = \lambda T z - 1 - \log \lambda T z 
\eeq
The Cramer theorem then states that 
\beq
\label{Cramer}
P \left(  \frac{Z_1 + \ldots + Z_{n}}{n} < \frac{1}{n}\right) \sim e^{ - n I(1)} = 
e^{ - n \left( \lambda T - \log \lambda T - 1 \right) }
\eeq


\begin{thebibliography}{99}
\bibitem{BK} K.~B{\"o}cker and C. Kl{\"o}ppelberg, "Operational VAR: a closed-form approximation",
RISK, December 2005, 90-93.
\bibitem{Clauset_2007} A.~Clauset, C. R.~Shalizi, and M.E.J.~Newman, "Power-law distributions in empirical data" (2007), available at http://arxiv.org/pdf/0706.1062.pdf.
\bibitem{Corless} R.M.~Corless, G.H.~Gonnet, D.E.G.~Hare, D.J.~Jeffrey, and D.E.~Knuth, "On 
the Lambert W function", Advances in Computational Mathematics {\bf 5} (1996), 329�359. 
\bibitem{IH_saddle} I.~Halperin, ``CDO Pricing with Saddle Point Method'', JPM (2005).

\bibitem{Hernandez} L.~Hernandez, J.~Tejero, A.~Suárez, and S.~Carrillo-Menéndez 
("Percentiles of Sums of Heavy-Tailed Random Variables: Beyond the Single-Loss Approximation",
Statistics and Computing, {\bf 24}(3) (2014), pp. 377-397, available at 
http://arxiv.org/pdf/1203.2564.pdf.
\bibitem{Landau} L.D.~Landau and E.M.~Lifshitz, "Quantum Mechanics", 
Butterworth-Heinemann (1981).
\bibitem{Newman_2006} M.E.J.~Newman, "Power Laws, Pareto Distributions and 
Zipf's Law", http://arxiv.org/pdf/cond-mat/0412004.pdf.

\bibitem{LR} R.~Lugannani and S.~Rice, "Saddlepoint approximations for the distribution of the sum of independent random variables", Advances in Applied Probability, v.12 (1980), pp. 475- 490.
\bibitem{MW} J.~Mathews and R.L.~Walker, {\it Mathematical Methods of Physics}, W.A. Benjamin, 
1964.
\bibitem{Sahay} A.~Sahay, Z.~Wan, and B.~Keller, "Operational risk capital: asymptotics in the 
case of heavy-tailed severity", {\it Journal of Operational Risk}, v.2 (2007), 61-72.
\bibitem{Tella} C.~Tellambura and D.~Senaratne, "Accurate Computation of the MGF of the 
Lognormal Distribution and its Application to Sum of Lognormals", {\it IEE Trans. on 
Communications}, {\bf 58} (2010), 1568-1577.
\bibitem{Varadhan} S.R.S. Varadhan, {\it Large Deviations and Applications}, SIAM, Philadelphia 
(1984).
\bibitem{ZJ} J.~Zinn-Justin, " The Principles of Instanton Calculus: a Few Applications", in {\it Recent
Advances in Field Theory and Statistical Mechanics}, Les Houches XXXIX (1982), ed. J.B.~Zuber and R.~Stora, Elsevier Science Publishers (1984).
\end{thebibliography}
\end{document}